\documentclass[a4paper,12pt]{article}
\usepackage{graphicx}
\usepackage{amssymb,amsmath}

\def\,{\ifmmode\mskip\thinmuskip\else\leavevmode\thinspace\fi}
\newcommand{\vecc}[1]{\mathbf{#1}}
\newcommand{\vecs}[1]{\mathbf{#1}{\lower-.2em\hbox{}^{2}}}

\newcommand{\be}{\begin{equation}}
\newcommand{\ee}{\end{equation}}
\newcommand{\ba}{\begin{eqnarray}}
\newcommand{\ea}{\end{eqnarray}}
\newcommand{\nn}{\nonumber}

\title{Process $ 3\rightarrow 3$ and crossing symmetry violation}
\date{}
\author {E.~Barto\v s$^{1,2}$, S.~R.~Gevorkyan$^{1,3}$, E.~A.~Kuraev$^1$}

\begin{document}
\maketitle
\begin{center} {
$^{1}$ \it Joint Institute of Nuclear Research, 141980 Dubna, Russia\\
$^{2}$ \it Dept. of Theor. Physics, Comenius Univ., 84248
Bratislava, Slovakia \\
$^{3}$ \it  Yerevan Physics Institute, 375036 Yerevan, Armenia }
\end{center}

\vspace*{2cm}

\begin{abstract}

Using the Sudakov technique we sum the perturbation series for the
process $3\to 3$ and obtain the compact analytical expression for
the amplitude of this process, which takes into account all
possible Coulomb interactions between colliding particles. Compare
it with the amplitude of the lepton pair production in heavy ion
collision i.e. in the process $ 2\to 4$, we show that crossing
symmetry between this processes holds only if one neglects the
interaction of produced pair with ions (i.e. in the approximation
$Z_{1,2}\alpha \ll 1$).
\end{abstract}

\section{Introduction}
During the past decade the growing interest to the process of
lepton pair production in the strong Coulomb fields appeared. This
is connected mainly with beginning of operation the relativistic
heavy ion collider RHIC (Lorentz factor $\gamma=\frac{E}{m}=108 $)
and the new collider LHC ($ \gamma=3000 $) which will operate in
the nearest future. At such energies the lepton pairs yield
becomes huge (remind that the total cross section of the process
$A+B\to A+B+e^++e^-$ grows with energy as the third power of the
logarithm \cite{LL,R}), so that a detail analysis of this process,
with taking into account the Coulomb corrections (CC)  is
required. Such work has been done during last years and a lot of
papers are devoted to this subject. Nevertheless the problem turn
out to be more complex than it seems from the first glance. For a
review of the works in this direction see for example \cite{LMS}
and references therein. We want only to notice the exciting result
obtained in the series of papers \cite{BM}: the Coulomb
corrections to the process $ A+B\to A+B+e^++e^-$ enters the
amplitude of this process in such a way that its  cross section is
determined solely by lowest (Born) term. At present we understand
that this result is the incorrect application of crossing symmetry
property which, as it is known long ago (see e.g. \cite{KLS}), is
valid only on the Born level. As an obvious example of the
crossing symmetry violation we want to cite the process of lepton
pair photoproduction on the nuclei and its counterpart  the
bremsstrahlung in lepton--nucleus scattering. Amplitudes of both
processes are determined by Coulomb phase which is infrared stable
in the case of pair photoproduction while it is infrared divergent
in the case of bremsstrahlung and this difference can't be adjust
by trivial crossing change of variables. Nevertheless taking into
account the importance of the problem and permanent interest to it
from scientific society we calculate the full amplitude for the
process $ 3\to 3$ accounting all possible photon exchanges among
the colliding relativistic particles. From its comparison with the
amplitude of the process $2\to 4$ one can see that the crossing
symmetry property becomes invalid whereas one takes into account
the final state interaction.

\section{ The Born amplitude}

\begin{figure}[thb]
\centering
\includegraphics[scale=.8]{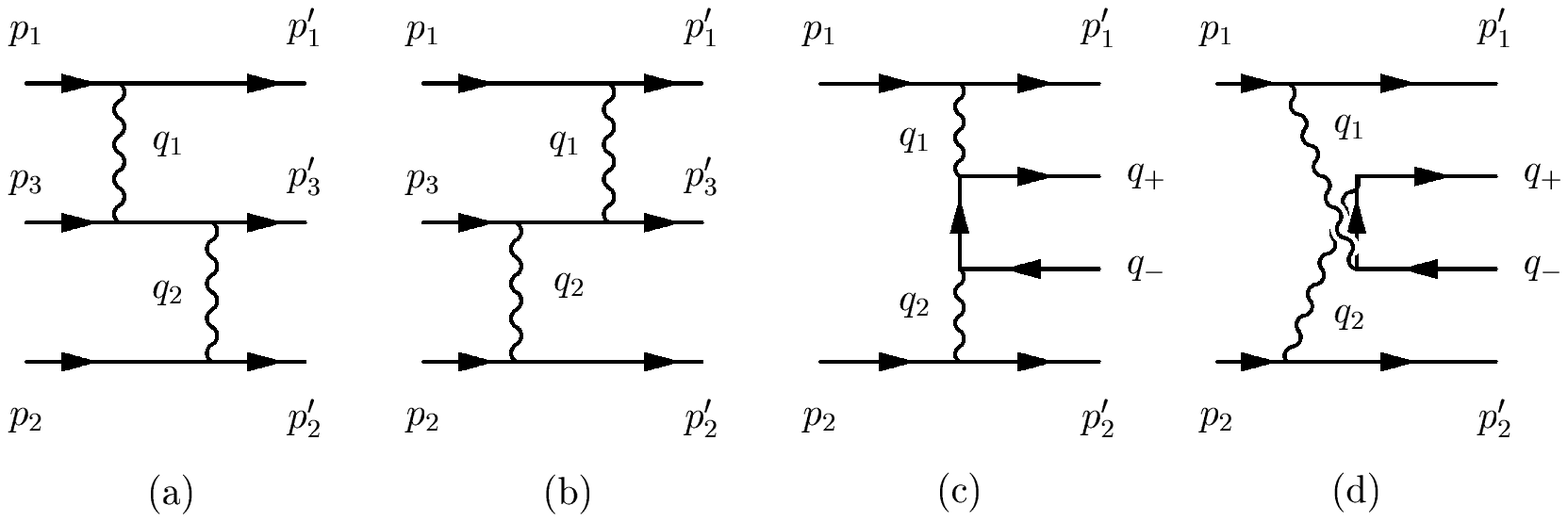}
\caption{Feynman diagrams for Born amplitudes of the process
$A_1+A_2+C\to A_1+A_2+C$ (a, b) and the process $A_1+A_2\to
A_1+A_2+C+\bar{C}$ (c, d).} \label{fig:1}
\end{figure}

Let us construct the amplitude of the process  $3\to 3$
represented in Fig.~\ref{fig:1} (a, b)
\begin{equation}\label{eq:1}
A_1(p_1)+A_2(p_2)+C(p_3)\to A_1(p_1')+A_2(p_2')+C(p_3').
\end{equation}
We consider the kinematics when all the energy invariants which
determined the process (\ref{eq:1}) are large, compared with the
masses of involved particles and the transferred momenta
\begin{gather}
s=(p_1+p_2)^2,\quad s_1=(p_1+p_3)^2,\quad s_2=(p_3+p_2)^2, \nn\\
 q_1^2=(p_1-p_1')^2,\quad q_2^2=(p_2-p_2')^2,\quad
q_3^2=(p_3-p_3')^2, \\ \nn p_1^2={p_1'}^2=m_1^2,\quad
p_2^2={p_2'}^2=m_2^2,\quad p_3^2={p_3'}^2=m^2,\\ \nn s\gg s_1 \sim
s_2\gg-q_1^2 \sim -q_2^2 \sim -q_3^2.
\end{gather}
\begin{figure}[thb]
\centering
\includegraphics[scale=.8]{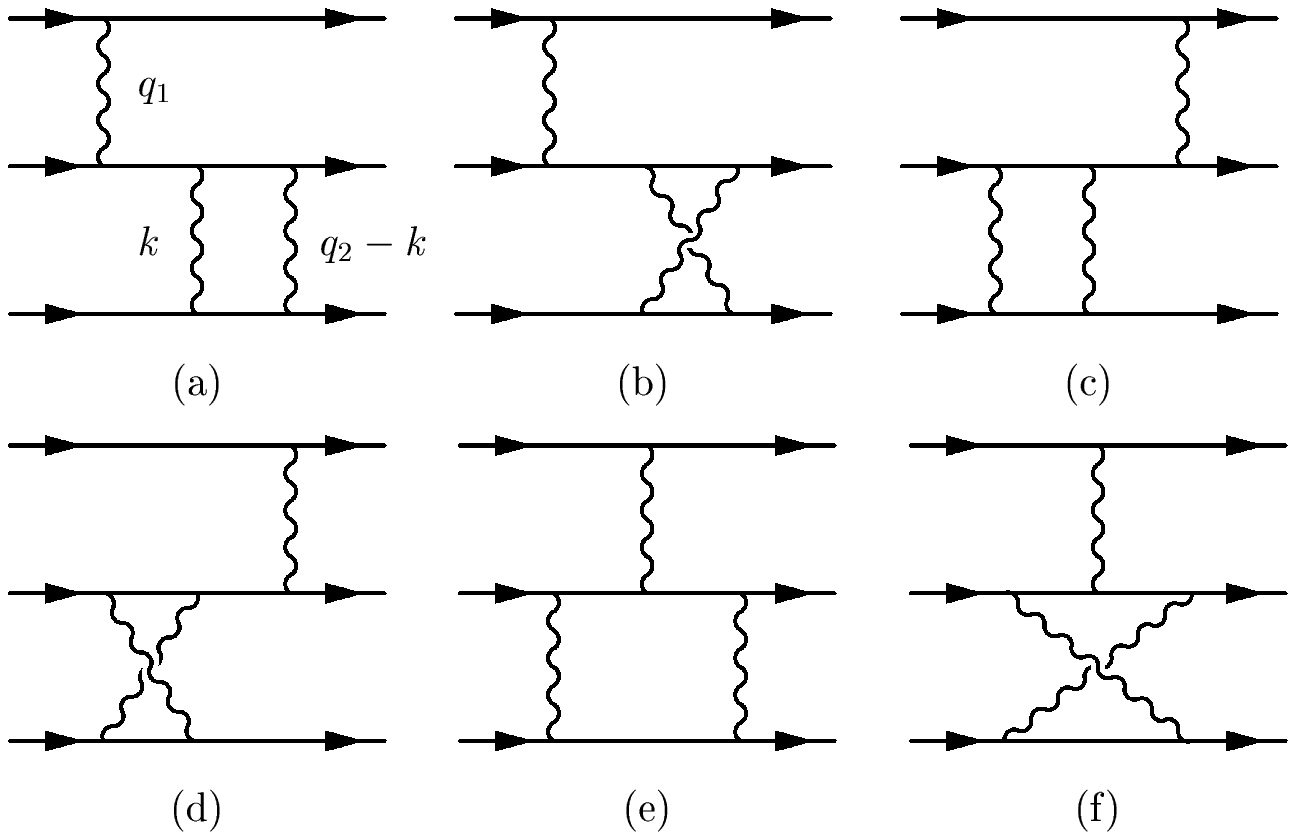}
\caption{Feynman diagrams for the process $A_1+A_2+C\to A_1+A_2+C$
with three photon exchange.} \label{fig:2}
\end{figure}
For the Born amplitude of the process (\ref{eq:1}) one can write
\begin{eqnarray}M_{(1)}^{(1)}=-i(4\pi\alpha)^2 Z_1 Z_2
 \bar{u}(p_1')\gamma_\mu u(p_1)\bar{u}(p_2')\gamma_\nu u(p_2)
\frac{\bar{u}(p_3') O_{\rho\sigma} u(p_3)
g^{\mu\sigma}g^{\nu\rho}}{q_1^2 q_2^2},
\end{eqnarray}
where $Z_{1,2}$ are the charge numbers of the colliding nuclei. We
use Sudakov's parameterization for all four--momenta entering  the
problem (for details see \cite{BGKN1})
\begin{gather}
q_1=\alpha_1 \tilde{p}_2+\beta_1\tilde{p}_1+q_{1\bot},\quad
q_2=\alpha_2 \tilde{p}_2+\beta_2\tilde{p}_1+q_{2\bot},  \nn\\
p_1'=\alpha_1'\tilde{p}_2+\beta_1'\tilde{p}_1+p_{1\bot}',\quad
p_2'=\alpha_2' \tilde{p}_1+\beta_2'\tilde{p}_1+p_{2\bot}', \\
p_3=\alpha_3 \tilde{p}_1+\beta_3\tilde{p}_1+p_{3\bot},\quad
p_3'=\alpha_3'\tilde{p}_1+\beta_3'\tilde{p}_1+p'_{3\bot},\nonumber
\end{gather}
and the Gribov's decomposition of the metric tensor in the
longitudinal and transverse parts
\[
g_{\mu\nu}=g_{\bot\mu\nu}+\frac{2}{s}\left(\tilde{p}_{1\mu}
\tilde{p}_{2\nu}+\tilde{p}_{1\nu}\tilde{p}_{2\mu}\right),
\]
with light--like 4-vectors $\tilde{p}_{1,2}$. For the kinematics
of the process we have
\begin{eqnarray}&&s=2\tilde{p}_1\tilde{p}_2,\quad \beta_1+\beta_3=\beta_3',\quad
\alpha_2+\alpha_3=\alpha_3', \\ \nn &&\quad \quad
g_{\mu\sigma}=\frac{2}{s}\tilde{p}_{1\sigma}\tilde{p}_{2\mu},\quad
g_{\nu\rho}=\frac{2}{s}\tilde{p}_{1\nu}\tilde{p}_{2\rho}, \\ \nn
&&\quad q_1^2=q_{1\bot}^2=-\vecs{q_1},\quad
q_2^2=q_{2\bot}^2=-\vecs{q_2},
\end{eqnarray}
where we use two--dimensional vectors in the plane transverse to
the z-axes, which we choose along 3--vector $\vec p_1=-\vec p_2 $
in the center of mass frame of initial particles $A_1$, $A_2$.
Using the gauge invariance conditions
\begin{eqnarray}\nonumber q_{1\rho}\bar{u}(p_3') O_{\rho\sigma} u(p_3)\approx
(\beta_1\tilde{p}_1+q_{1\bot})_\rho \bar{u}(p_3') O_{\rho\sigma}
u(p_3)=0,\\ q_{2\sigma}\bar{u}(p_3') O_{\rho\sigma} u(p_3)\approx
(\alpha_2\tilde{p}_2+q_{2\bot})_\sigma \bar{u}(p_3')
O_{\rho\sigma} u(p_3)=0,
\end{eqnarray}
we get the Born amplitude in the form
\begin{eqnarray}\label{eq:3}
M_{(1)}^{(1)}(q_1,q_2)=-4 i s N_1 N_2(4\pi\alpha Z_1)(4\pi\alpha
Z_2)B(q_1,q_2),
\end{eqnarray}
with
\begin{eqnarray}& \quad\quad B(q_1,q_2)=\displaystyle\frac{\bar{u}(p_3')
O_{\rho\sigma} u(p_3) q_{1\bot}^\rho q_{2\bot}^\sigma
}{s\alpha_2\beta_1 \vecs{q_1}\vecs{q_2}},& \\ \nn & \displaystyle
N_1=\frac{1}{s}\bar{u}(p_1')\hat{p}_2u(p_1),\quad
N_2=\frac{1}{s}\bar{u}(p_2')\hat{p}_1u(p_2),&  \\ \nn &
s\alpha_2\beta_1=-q_3^2-(\vecc{q_1}-\vecc{q_2})^2\sim m^2.&
\end{eqnarray}
The values of $N_i$ for every polarization state of initial
particles (or for spinless particles) are unity and
\begin{eqnarray}&&\bar{u}(p_3') O_{\rho\sigma} u(p_3)q_{1\bot}^\rho
q_{2\bot}^\sigma=\\
&&\quad\bar{u}(p_3')\left[\hat{q}_{2\bot}\frac{\hat{p}_3+\hat{q}_1+m}
{(p_3+q_1)^2-m^2}\hat{q}_{1\bot}+
\hat{q}_{1\bot}\frac{\hat{p}_3+\hat{q}_2+m}{(p_3+q_2)^2-m^2}
\hat{q}_{2\bot}\right]u(p_3). \nonumber
\end{eqnarray}
\begin{figure}[tb]
\centering
\includegraphics[scale=.8]{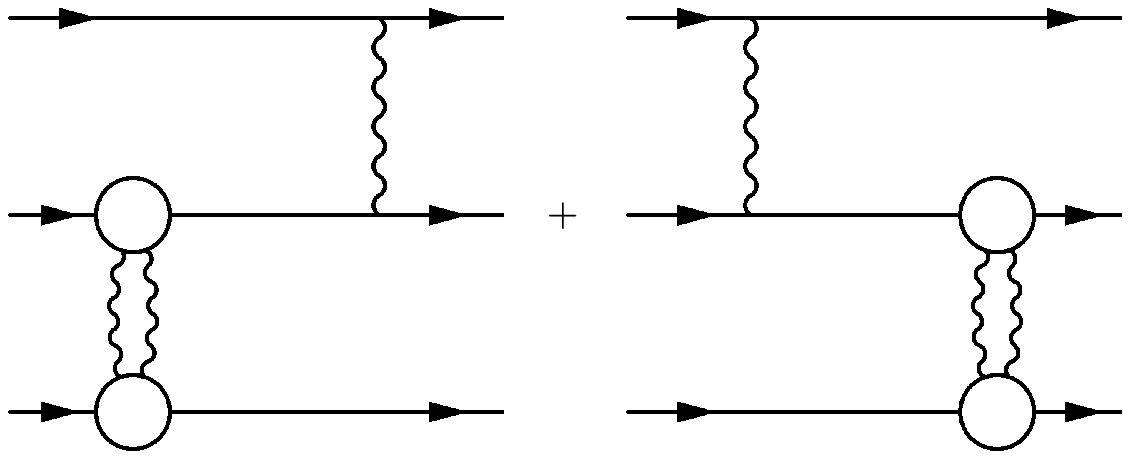}
\caption{Feynman diagram for the amplitude $M_{(2)}^{(1)}$.}
\label{fig:3}
\end{figure}

\section{The Coulomb corrections to the process $3\to 3$}

Let us consider the set of six Feynman diagrams (FD) with one
virtual photon connected the $p_3$ line with the particle $A_1$
and the two ones connected $p_3$ line with the particle $A_2$ (see
Fig.~\ref{fig:2}). The loop momentum integration in the relevant
matrix element can be performed accounting that
\begin{equation}
d^4 k=(2\pi i)^2 \frac{1}{2s}\frac{d(s\alpha_k)}{2\pi i}
\frac{d (s\beta_k)}{2\pi i}d^2k_\bot, \quad
k=\alpha_k\tilde{p}_2+\beta_k\tilde{p}_1+k_{\bot}.
\end{equation}

It can be shown that only 4 FD amplitudes works (Fig.~\ref{fig:2}
(a--d)). Really, when one performs the $\alpha_k$ integration in
Fig.~\ref{fig:2}~(e, f), one has a situation when both poles in
$\alpha_k$ complex plane are situated in the same half--plane, so
their contribution to the amplitude is zero (suppressed by factor
$|q_3^2/s|\sim |s_1/s|$).

The physical reason of this suppression is the same as in the case
of bremsstrahlung suppression for fast charged particle moving
through the media known as the Landau--Pome\-ran\-chuk effect.
Really, this effect can be explained starting from the fact of
power suppression of radiation between two scattering centers in
the case when the distance between these centers is less than the
coherent length.

It is convenient to introduce 8 FD (including remaining four ones
and additional four FD with interchanged photons absorbed by
nucleus $A_2$ line). To avoid the double counting we multiply
 the relevant matrix element by statistical factor $1/2!$. This
 trick permits  to perform the integration over Sudakov variables
 $\alpha_k$, $\beta_k$ with the result
\begin{multline}
\int\frac{d(s\alpha_k)}{2\pi i}\left(\frac{1}{s\alpha_k+i0}+
\frac{1}{-s\alpha_k+i0}\right)=\\   \int\frac{d(s\beta_k)}{2\pi i}
\left(\frac{1}{s\beta_k+i0}+\frac{1}{-s\beta_k+i0} \right)=1.
\end{multline}
Further integration over transverse momentum is straightforward
\begin{equation}\int\frac{d^2\vecc{k}}{\pi}\frac{1}{(\vecc{k}+\lambda)
((\vecc{q_2}-\vecc{k})^2+\lambda^2)}=\frac{2}{\vecs{q_2}}
\ln\frac{\vecs{q_2}}{\lambda^2}.
\end{equation}\begin{figure}[tb]
\centering
\includegraphics[scale=.78]{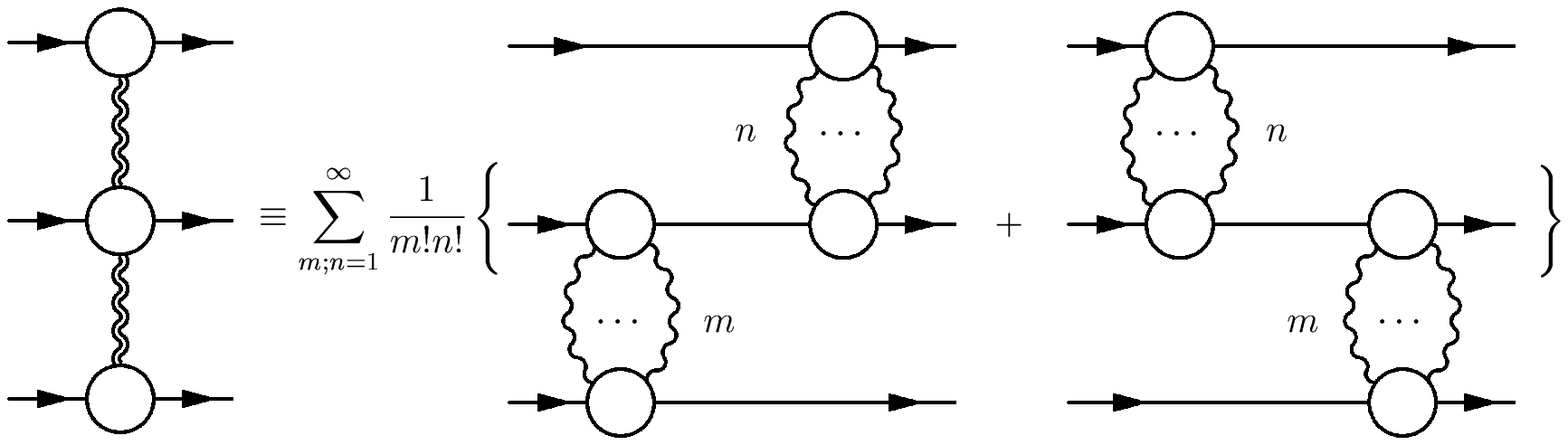}
\caption{Feynman diagram for the amplitude $M_{(\infty)}^{
(\infty)}(q_1,q_2)$.} \label{fig:4}
\end{figure}

For the amplitude $M_{(2)}^{(1)}$ (see Fig.~\ref{fig:3}) and the
similar amplitude $M_{(1)}^{(2)}$ we obtained
\begin{equation}\label{eq:13}
M_{(2)}^{(1)}+M_{(1)}^{(2)}=M_{(1)}^{(1)}(q_1,q_2)\frac{1}{2!}\left[
2iZ_1\alpha\ln\frac{\vecs{q_1}}{\lambda^2}+2iZ_2\alpha\ln\frac{
\vecs{q_2}}{\lambda^2}\right].
\end{equation}The amplitude for arbitrary amount of interchanged photons (see
Fig.~\ref{fig:4}) is constructed in the  similar way
\begin{equation}\label{eq:14}
M_{(\infty)}^{(\infty)}(q_1,q_2)=M_{(1)}^{(1)}(q_1,q_2)
e^{i(\varphi_1(\vecc{q_1})+\varphi_2(\vecc{q_2}))}
\end{equation}with the Coulomb phases
\[
\varphi_1(\vecc{q_1})=Z_1\alpha\ln\frac{\vecs{q_1}}{\lambda^2},\quad
\varphi_2(\vecc{q_2})=Z_2\alpha\ln\frac{\vecs{q_2}}{\lambda^2}.
\]

Consider now the case with one additional exchanged photon between
two nuclei $A_1$, $A_2$. The relevant matrix element $M_{(1B)}$
reads
\begin{eqnarray}  \label{eq:15}
M_{(1B)}=i\alpha Z_1Z_2\int\frac{d^2\vecc{k}}
{\pi(\vecc{k}^2+\lambda^2)}M_{(1)}^{(1)}(q_1+k,q_2+k).
\end{eqnarray}
Two--dimensional integral in (\ref{eq:15}) is infrared divergent.
To regularize it we introduce the photon mass parameter $\lambda$.

\begin{figure}[tb]
\centering
\includegraphics[scale=.8]{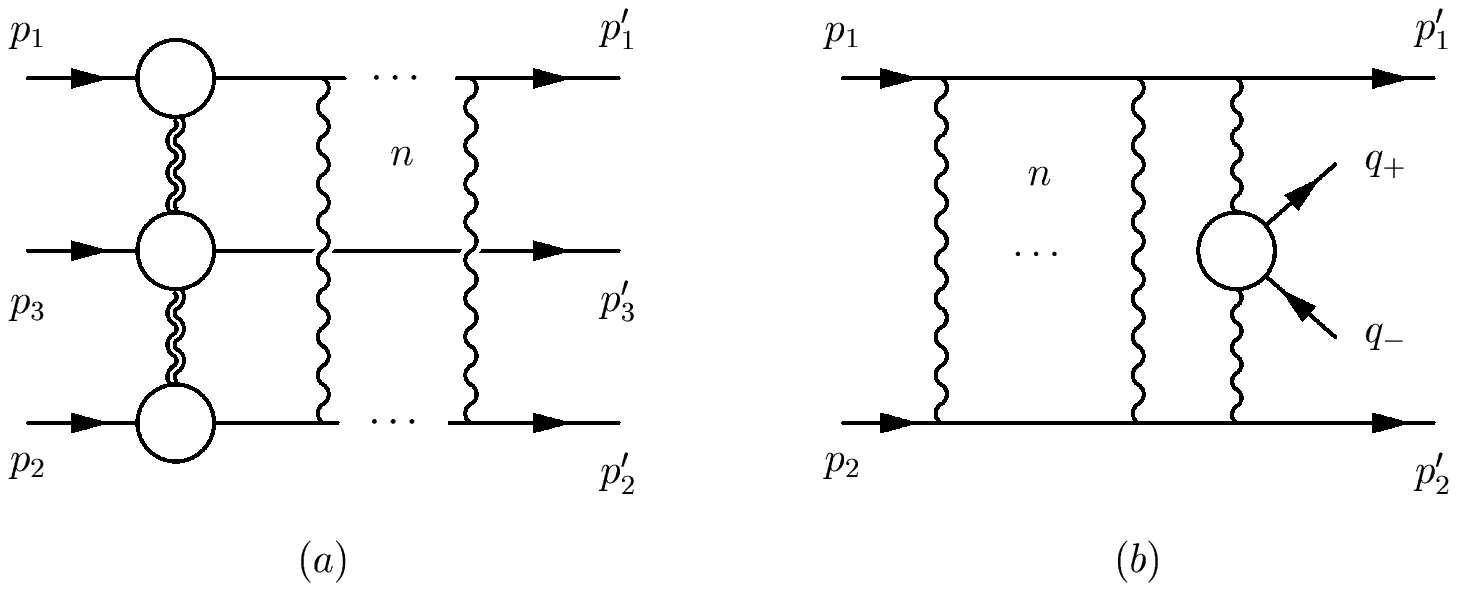}
\caption{Feynman diagrams for the $n$ photon exchange between
nuclei $A_1$ and $A_2$ compared with the Born diagram for the
processes $3\rightarrow 3$ (a) and $2\rightarrow 4$ (b) (blob in
(b) correspond to diagrams in Fig.~\ref{fig:1} (c, d)) .}
\label{fig:5}
\end{figure}

In the same approach we get for the matrix element with the $n$
exchanged (between nuclei) photons (see Fig.~\ref{fig:5} (a))
\begin{equation}M_{(nB)}=\frac{(i\alpha Z_1Z_2)^n}{
n!}\prod_{i=1}^n\int\frac{d^2\vecc{k_i}}
{\pi(\vecs{k_i}+\lambda^2)} M_{(1)}^{(1)}(q_1+\sum_{i=1}^n
k_i,q_2+\sum_{i=1}^n k_i).
\end{equation}
It is convenient to write down this expression in impact parameter
representation. For this aim we use the following identity
\begin{multline}   \label{eq:17} \int
d^2\vecc{k_{n+1}}\delta^{(2)}(k_{n+1}-q_1-\sum_{i=1}^n k_i)=\\
\frac{1}{(2\pi)^2} \int d^2\vecc{k_{n+1}}d^2 \boldsymbol\rho
e^{i(\vecc{k_{n+1}}-\vecc{q_1}-\sum \vecc{k_1})\cdot
\boldsymbol\rho}=1.
\end{multline}
Thus the matrix element with arbitrary number of exchanged photons
can be cast
\begin{equation}\label{eq:18}
M(3\rightarrow 3)=\sum_{n=1}^\infty M_{(nB)}=\frac{1}{4}
\int\frac{d^2 \boldsymbol\rho} {\pi}
e^{-i\vecc{q_1}.\boldsymbol\rho} e^{i\alpha
Z_1Z_2\psi(\boldsymbol\rho)}\tilde{M}_{(1)}^{(1)}(\rho, q_1, q_2)
\end{equation}with
\begin{eqnarray}\psi(\boldsymbol\rho)=\int\frac{d^2\vecc{k}}{\pi}
\frac{e^{-i\vecc{k}.\boldsymbol\rho}}{\vecc{k}^2+\lambda^2}=
2K_0(\rho\lambda)=-2(\ln \rho m+\ln\frac{\lambda}{m}-\ln2+C),
\end{eqnarray}
where $C=0.577$ and
\begin{equation}\label{eq:20}
\tilde{M}_{(1)}^{(1)}(\rho,q_1,q_2)=\int\frac{d^2\vecc{k}}{\pi}
e^{-i\vecc{k}.\boldsymbol\rho}M_{(1)}^{(1)}(k,k+q_2-q_1).
\end{equation}

This result confirms the general ansatz given above (see
(\ref{eq:14})) that all the dependence on "photon mass" $\lambda$
can be represented as a phase factor. As can be seen the whole
amplitude (\ref{eq:18}) cannot be cast solely as a Born amplitude
multiplied by the phase factor. The corresponding contributions to
the total cross section (except the Born term) will be enhanced
only by the first power of logarithm in energy.

Finally, taking into account all photon exchanges between particle
$C$ and nuclei $A_1$, $A_2$ we obtain the general answer by the
simple replacement in the expression (\ref{eq:20})
\begin{equation}
M_{(1)}^{(1)}\rightarrow M_{(\infty)}^{(\infty)}=M_{(1)}^{(1)}
(k,k+q_2-q_1) e^{i\varphi_1(\vecc{k})+
i\varphi_2(\vecc{k}+\vecc{q_2}-\vecc{q_1})}
\end{equation}
with $\varphi_1$, $\varphi_2$ given in (\ref{eq:14}).

\section{The Coulomb corrections to the process $2\to 4$}

As was mentioned above, our goal is to investigate the crossing
symmetry property between the amplitudes of the process
(\ref{eq:1}) and the relevant process in Fig.~\ref{fig:1} (c, d)
\begin{equation}\label{eq:22}
A_1(p_1)+A_2(p_2)\to A_1(p_1')+A_2(p_2')+C(p_3)+\bar{C}(p_4),
\end{equation}
with the following kinematics
\begin{gather}
s=(p_1+p_2)^2,\quad s_{p}=(q_++q_-)^2,\nn\\
q_1^2=(p_1-p_1')^2,\quad q_2^2=(p_2-p_2')^2,\\ \nn
p_1^2={p_1'}^{2}=m_1^2,\quad q_+^2=q_-^2=m_2^2,\quad
q_+^2=q_-^{2}=m^2,\\ s \gg-q_1^2 \sim -q_2^2 \sim s_{12}.\nn
\end{gather}

Using the Sudakov technique the Born amplitude for the process
(22) can be represented in the form
\begin{align} \label{eq:24}
M_p&= -is2^6\pi^2Z_1Z_2N_1N_2B_p(q_1,q_2)
\end{align}
with
\begin{gather*}
B_p(q_1,q_2)=\frac{e_1^\alpha e_2^\beta
\bar{u}(q_-) T_{\alpha\beta}v(q_+)|\vecc{q_1}|
|\vecc{q_2}|}{\tilde s\vecs{q_1}\vecs{q_2}}, \\\tilde
s=s\alpha_2\beta_1=(q_++q_-)^2+
(\vecc{q_1}+\vecc{q_2})^2,\\
T_{\alpha\beta}=\gamma_\beta \frac{{\hat q_1-\hat
q_++m}}{{(q_1-q_+)^2-m^2}}\gamma_\alpha+\gamma_\alpha \frac{{\hat
q_2-\hat q_++m}}{{(q_2-q_+)^2-m^2}}\gamma_\beta.
\end{gather*}
(for details see \cite{BGKN1}).

Generalization for the case of arbitrary number of exchanged
photons between colliding nuclei is straightforward. The same
approach as above leads to the following form of the generalized
amplitude
\begin{eqnarray}\label{eq:25}
M(2\rightarrow 4 )=\frac{1}{4}\int
\frac{d^2{\boldsymbol\rho}}{\pi}
e^{-i\vecc{q_1}{\boldsymbol\rho}-i\alpha
Z_1Z_2\psi({\boldsymbol\rho})}\Phi_B({\rho},{q_2}),
\end{eqnarray}
with
\begin{gather*}
\Phi_B(\rho,{q_2})=\int\frac{d^2\vecc{k}}
{\pi}e^{i\vecc{k}\boldsymbol\rho} M_p({k},{q_2}-{k}).
\end{gather*}

Comparing the expression (\ref{eq:25}) with the amplitude for the
process $3\rightarrow 3$ (\ref{eq:18}) one can see that the
crossing symmetry property between the considered processes takes
place in the case when one neglects the multiple exchanges of
particle $C$ with nuclei. Moreover, this statement is correct even
when one takes into account the screening effects between nuclei
$A_1$ and $A_2$ in both processes, which manifest itself by
insertion of light--by--light scattering blocks into Feynman
amplitudes. As was shown in \cite{BGKN2} accounting of this effect
can be provided by the universal factor
\begin{equation}
\exp{\left\{-\frac{\alpha^2Z_1Z_2}{2}LA(\rho)\right\}},\quad
L=\ln(\gamma_1\gamma_2),
\end{equation}
with the complex quantity $A(\rho)$ connected with the Fourier
transformation of light--by--light scattering amplitude and
$\gamma_{1}$, $\gamma_{2}$ the Lorentz factors of colliding
nuclei.

Nevertheless, crossing symmetry is broken in all orders of
perturbation theory if one try to compare the full amplitude for
the process $3\to 3$ (expression (24) with the replacement (21))
and the relevant amplitude for the process $2\to 4$ accounting the
multiple interaction of produced particles \cite{BGKN1}.

Thus the crossing symmetry property takes place only for the
colliding nuclei with charge numbers fulfilled the approximation
$Z_{1,2}\alpha\ll 1$.

\section*{Acknowledgment}
We are grateful to V.~Serbo and N.~Nikolaev for the useful
discussions and comments. The work is supported by grants INTAS
97-30494, SR-2000.


\begin{thebibliography}{99}
%
\bibitem{LL} L.~D.~Landau and E.~M.~Lifshits, Phys.~Z.~Sowjet.
6 (1934) 244.
%
\bibitem{R} G.~Racah, Nuovo Cimento 14 (1937) 93.
%
\bibitem{LMS} R.~N.~Lee, A.~I.~Milstein, V.~G.~Serbo, hep-ph/0108014.
%
\bibitem{BM}
A.~J.~Baltz, L.~McLerran, Phys.~Rev.~C58, (1998) 1679.
%
\bibitem{KLS}
E.~A.~Kuraev, L.~N.~Lipatov, M.~I.~Strikman, Soviet ZhETP 66
(1974) 838.
%
\bibitem{BGKN1}
E.~Barto\v{s}, S.~R.~Gevorkyan, E.~A.~Kuraev and N.~N.~Nikolaev,
hep-ph/0109281.
%
\bibitem{BGKN2}
E.~Barto\v{s}, S.~R.~Gevorkyan, E.~A.~Kuraev and N.~N.~Nikolaev,
hep-ph/0204327.
%
\bibitem{GPS}
I~Ginzburg, S.~Panfil, V.~Serbo, Nucl. Phys. B284 (1987) 685.
%
\end{thebibliography}
\end{document}